# A new Method for the in vivo identification of material properties of the human eye. Feasibility analysis based on synthetic data

**Stefan Muench**[1*], **Mike Roellig**[1], **Daniel Balzani**[2]

[1]Fraunhofer Institute for Ceramic Technologies and Systems IKTS, Department of Testing of Electronics and Optical Methods, Dresden, Germany

[2]Ruhr University Bochum, Chair of Continuum Mechanics, Bochum, Germany

**\* Correspondence:**
Daniel Balzani
daniel.balzani@rub.de

**Abstract**

This paper proposes a new method for in vivo and almost real-time identification of biomechanical properties of the human cornea based on non-contact tonometer data. Further goal is to demonstrate the method's functionality based on synthetic data serving as reference. For this purpose, a finite element model of the human eye is constructed to synthetically generate displacement full-fields from different datasets with keratoconus-like degradations. Then, a new approach based on the equilibrium gap method (EGM) combined with a mechanical morphing approach is proposed and used to identify the material parameters from virtual test data sets. In a further step, random absolute noise is added to the virtual test data to investigate the sensitivity of the new approach to noise. As a result, the proposed method shows a relevant accuracy in identifying material parameters based on displacement full fields. At the same time, the method turns out to work almost in real-time (order of a few minutes on a regular work station) and is thus much faster than inverse problems solved by typical forward approaches. On the other hand, the method shows a noticeable sensitivity to rather small noise amplitudes. However, analysis show that the accuracy is sufficient for the identification of diseased tissue properties.

## 1    Introduction

Eye diseases are particularly harmful for those affected. They impair vision and thus separate people from their environment and society. Especially children and adolescents are handicapped in their individual development by eye diseases. A typical eye disease that occurs during puberty is keratoconus [1–3]. The disease results in softening and deformation of the transparent anterior part of the outer shell of the eye, called cornea, under intraocular pressure (IOP) [2–4]. This leads to defective vision. A fast, early and reliable diagnosis of eye diseases such as keratoconus is essential for early treatment. The objective is to keep the impact on vision to a minimum. Measuring the biomechanical properties of the cornea in vivo is promising for this purpose. So far, they cannot be measured non-destructively. On the contrary, the lack of knowledge about the biomechanical properties and their neglect in the measurement of other quantities, such as the IOP, leads to measurement deviations and incorrect estimations of the IOP [5].

Air pulse-based non-contact tonometers (NCT) such as the Ocular Response Analyzer (ORA) from Reichert Inc. (Buffalo, USA) and the Corvis® ST from Oculus Optikgeräte GmbH (Wetzlar, Germany) are widely used as standard methods for intraocular pressure measurement. Especially during the measurement with the Corvis® ST, large amounts of load-displacement data are generated, which have not been fully utilized so far. Currently, two stiffness parameters SP-A1 and SP-HC exist, which indicate the resistance of the cornea to deformation. They are calculated from pressures determined by the device at specific times with respect to the indentation depth of the cornea [6, 7]. In addition, the Corvis biomechanical index (CBI) is used. It is based on a nonlinear function of various Corvis® ST parameters determined via a regression analysis to separate healthy eyes from eyes suffering from keratoconus [7]. Another new approach is the stress-strain index (SSI). It reduces the comparison of a measured stress-strain curve with a reference curve for healthy corneas to a scalar value. The reference curve is based on experimental data [8]. The mentioned quantities are based on purely empirical functions, so that they are only applicable in defined ranges. For example, the SSI is only valid for corneas with normal topography, which already excludes keratoconic eyes [8]. Furthermore, the parameters do not provide structural information of the corneal tissue. However, the available data provide the basis for linking biomechanical properties of the cornea with structural material properties such as collagen fiber stiffness. Such parameters obtained in vivo would be a key to earlier and more accurate diagnosis of many other eye diseases. Consequently, there are already publications to use the experimental basis from the NCT for the inverse identification of biomechanical material properties using finite elements (FE) [9, 10]. In this process, associated mechanical fields inside the cornea are calculated based on specified deformations or forces and subsequently compared with the experimental data. However, these so-called finite element update methods (FEMU) are computationally expensive, and so it is very difficult to raise them out of the laboratory stage. Therefore, in the following we would like to propose a method that allows both a determination of structural material properties and to be update free and thus almost real-time capable. Since currently not all necessary input variables can be acquired by the Corvis® ST, an approach for data enrichment is presented, which forms the link between a realistic measurement and the presented inverse approach.

## 2    Materials and Methods

For the investigation of the method, synthetic data obtained from FE simulations of the NCT are used as a database. The advantage of this approach is that reference data sets with uniquely assigned material properties can be generated for test purposes. Thus, the accuracy of the presented approaches can be quantified and investigations regarding the influence of image noise on the determined characteristic values can be carried out. The FE simulations were performed with the software ANSYS® Mechanical APDL, Release 18.2.

**Virtual Non-Contact Tonometry**

In order to enable a realistic computational simulation of non-contact tonometry, an accurate model of the human eye is necessary. The eye shell consists of two dominant primary geometries, see Figure 1. In the proposed model, the cornea is represented as an ellipsoid. Its major semiaxis is coincident with the symmetry axis of the eye. The sclera is simplified to a sphere. Due to the symmetrical structure, the eye can be transferred into a quarter model. According to reality, the cornea has a lower thickness in the center than in the peripheral region. The limbus forms a 2 mm wide transition from the cornea to the sclera at about 4.6 mm. The thickness of the sclera is kept constant for simplification.

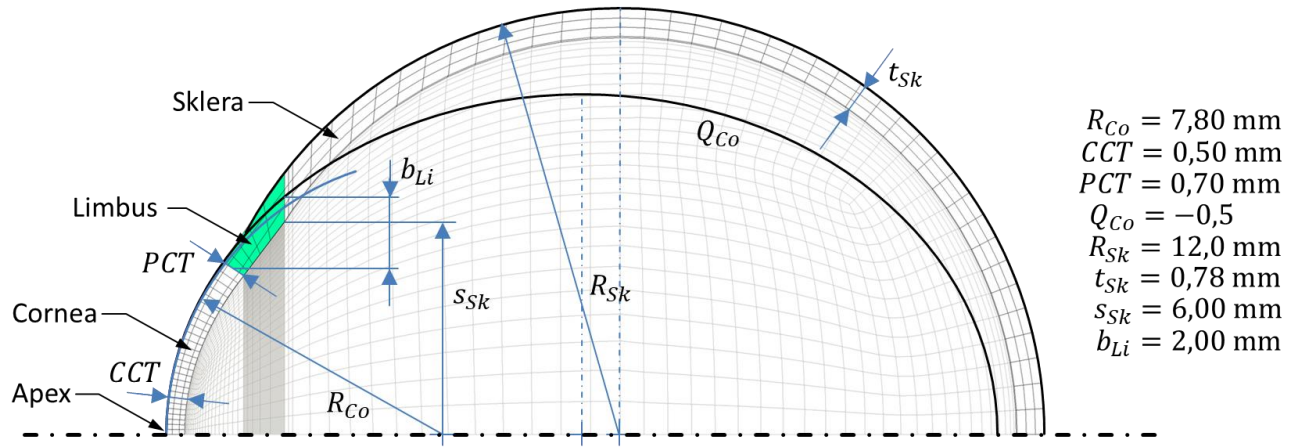

**Figure 1: FE model of the eye. Lens and hydrostatic fluid elements are hidden for clarity. The geometric parameters are based on values according to [11].**

For the process of NCT, the system response of the eye is important. This includes not only the deformation of the eye shell but also the varying IOP. The lens (1) separates the interior of the eye into two sections (2, 3), see Figure 2. This separation leads to different intraocular pressures in the sections during the NCT. Accordingly, the modelling of the interior of the eye is also of high importance. The lens is rotationally symmetrical to the symmetry axis of the eye. Its front side is flattened compared to the back side. For simplicity, the lens, zonula fibers and trabecular meshwork are modeled as one volume (1). In the model, it is suspended underneath the limbus. The iris is neglected with respect to a very small mechanical influence on the corneal deformation. The anterior and posterior chambers of the eye are combined into one volume (2). This is meshed with hydrostatic fluid elements, which calculate an IOP increase depending on the displaced volume. The vitreous body (3) is also meshed with hydrostatic fluid elements.

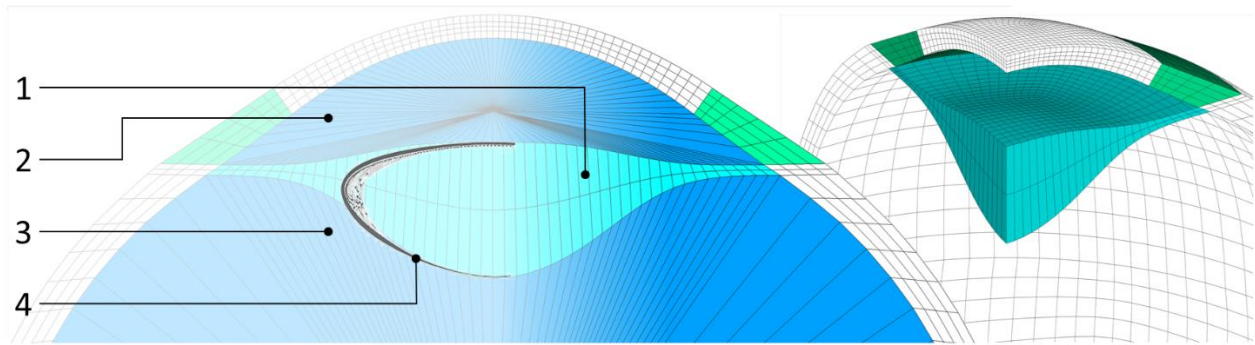

**Figure 2: Comparison of the lens (1) in the FE model with the typical lens geometry (4) according to [12]. The lens separates the interior of the eye into the anterior and posterior chamber volume (2) and the volume of the vitreous body (3).**

Mapped mesh technique has been used to discretize the geometry. Hexahedral 8-node elements represent the solid components: cornea, limbus, sclera and lens. In order to reduce the influence of the discretization on the results, a mesh influence study is performed according to [13]. The target parameters are the corneal deflection amplitude $Def A$, the distance of the bending maxima $PD$ and the $IOP_{AC}$ in the anterior chamber of the eye. They are determined at the time of the maximum pressure pulse applied to the cornea. For the applied mesh, grid convergence indices (GCI) of less than 1 % are obtained for all three target variables using 41184 solid elements and 10140 fluid elements.

To represent the complex material behavior of the cornea, consisting of extracellular matrix and collagen fibers, the model according to Sánchez, Moutsouris and Pandolfi [14] was used in a slightly modified manner and then implemented into ANSYS® using the USERMAT subroutine. It is assumed that the stroma has the main mechanical influence during NCT and that the remaining thin layers are negligible. Instead of representing them, they are added to the stroma and receive the same properties. Here, we only recapitulate the major components of the material model, which is mostly based on the assumption that there exists a hyperelastic strain energy density function of the structure

$$\psi = \psi_{\text{vol}}(J) + \tilde{\psi}_{\text{iso}}(\tilde{I}_1) + \sum_{i=1}^{n=2} \tilde{\psi}_{\text{ti},i}(\tilde{I}^*_{4,i}) \tag{1}$$

wherein the volumetric part $\psi_{\text{vol}}$ is used here as penalty function to account for quasi-incompressibility. The energy density $\psi_{\text{iso}}$ describes the response of the isotropic matrix material, the function $\psi_{\text{ti}}$ the response of the collagen fibers. As deformation measures, the determinant $J$ of the deformation gradient $\boldsymbol{F}$, the isochoric parts of the first principal invariant $\tilde{I}_1$ of the right Cauchy-Green tensor (in index notation for a fixated Cartesian coordinate system) $C_{jk} = F_{ij}F_{ik}$, and the isochoric part of the fourth modified invariant $\tilde{I}^*_4$ as a mixture of the fourth invariant $\tilde{I}_4$ defined in terms of $C_{jk}$ and the structural tensor $M_{jk} = a_j a_k$ with the fiber direction vector $\boldsymbol{a}$, and the invariant $\tilde{I}_1$:

$$\tilde{I}^*_4 = \kappa \tilde{I}_1 + (1 - 3\kappa)\,\tilde{I}_4 \tag{2}$$

with $\kappa \in [0, 1/3]$. The individual energy density functions are given by

$$\psi_{\text{vol}}(J) = \frac{1}{4} K\,[J^2 - 1 - 2\,ln(J)] \tag{3}$$

$$\tilde{\Psi}_{\text{iso}}(\tilde{I}_1) = \frac{\mu}{2}(\tilde{I}_1 - 3) \tag{4}$$

$$\tilde{\psi}_{\text{ti},i}(\tilde{I}^*_{4,i}) = -\frac{k_1}{2k_2} + \frac{k_1}{2k_2} exp\left[k_2(\tilde{I}^*_{4,i} - 1)^2\right] \underbrace{\left(1 + K^*_i \sigma^2_{\tilde{I}^*_{4,i}}\right)}_{a} \tag{5}$$

wherein $a$ is defined as the second order term of the Taylor expansion of $\tilde{\psi}_{\text{ti},i}$ around $\tilde{I}^*_4$. Based on the energy density $\psi$, the Cauchy stress tensor entering the equilibrium equation as basis of the structural boundary value problems considered later can be computed mainly as derivative of the energy with respect to the strains. The energy equations include the penalty parameter $K$, the components $\mu_1$ and $\mu_2$ of the shear modulus $\mu$ as well as the collagen fiber stiffness $k_1$ and the dimensionless material constant $k_2$ [14]. The anisotropic part of the energy density function is calculated from the sum of the transversely isotropic fiber components for mainly two collagen fiber directions $n = 2$. Represented are the two main directions nasal-temporal and superior-inferior with the location-dependent fiber dispersions corresponding to [15]. Since the difference between the first and second invariants $\tilde{I}_1$ and $\tilde{I}_2$ in the stretch range of the cornea during NCT is very small, this results in a low sensitivity with respect to the two associated parameters in the original model $\mu_1$ and $\mu_2$ and thus, a potentially overrated sensitivity in the inverse problem. For this reason, we propose the simplification of the Mooney-Rivlin model, which was used in the original model [13] for the isotropic response, to the Neo-Hooke model with only one parameter $\mu = \mu_1 + \mu_2$ shown in Eq. (4) for the application to the eye. One further argument for the change to the Neo-Hooke model becomes apparent when analyzing realistic parameters for $\mu_2$. Table 1 summarizes typical values from the literature for the parameters mentioned. They were inversely determined by the respective authors using different experimental data sets. As can be seen, in many cases negative values for $\mu_2$ are obtained rendering the isotropic energy function non-polyconvex and thus, not automatically materially stable [16]. However, not satisfying such generalized convexity conditions may allow for an unphysical response and thus, questionable numerical results. In the isotropic energy function, this problem is cured when the Neo-Hooke model is applied where only positive values of $\mu$ are obtained resulting in a polyconvex isotropic energy density. Since we focus on the model [13], which is a hyperelastic model, we assume that the deformations induced during NCT are still in a physiological regime and thus, not associated with microscopic damage and a resulting stress-softening response. If such supra-physiological deformation regimes are to be analyzed, convexified damage models [17] may be considered which have been also successfully applied for arterial walls with mainly two collagen reinforcements [18]. But for NCT such loading regimes should in fact be avoided rendering a hyperelastic formulation a reasonable choice. Based on the parameter data in Table 1, we define our reference data set for healthy corneas (last line in Table 1). For the investigation of the accuracy of the inverse method described later, degraded properties are also applied. A pathological change of the cornea in keratoconus leads to the reduction of collagen fiber stiffness [4]. Accordingly, three degraded material datasets KK-I, KK-II and KK-III with a reduced fiber stiffness value $k_1$ are defined. Table 2 summarizes the material parameters used during virtual NCT.

**Table 1: Inversely calculated material parameters from literature and calculation of the resulting parameter $\mu = \mu_1 + \mu_2$.**
**in - in vivo, ex - ex vivo, d - indentation, i - inflation, k - keratectomy, t - tensile test**

| Publication | $K$ [MPa] | $\mu_1$ [MPa] | $\mu_2$ [MPa] | $\mu$ [MPa] | $k_1$ [MPa] | $k_2$ [−] | exper. basis |
|---|---|---|---|---|---|---|---|
| Alastrué 2006 [19, 20] | 0.075 | 0.005 | 0 | 0.005 | 0.0049 | 314 | ex/i [21] |
| Pandolfi 2006 [20] | 5.500 | 1.055 | -1.0455 | 0.0005 | 0.250 | 50 | ex/t [22] |
|  |  | 7.555 | 7.550 | 0.005 | 0.175 | 15 | ex/t [23] |
|  |  | 0.00505 | -0.005 | 0.00005 | 0.055 | 16 | ex/t [24] |
|  |  | 0.055 | -0.005 | 0.050 | 0.055 | 14 | ex/t [25] |
| Petsche 2013 [26] | - | 0.00559 | - | 0.00559 | 0.638 | 314 | in/d |
| Sánchez 2014 [14] | 5.500 | 0.060 | -0.010 | 0.050 | 0.040 | 200 | in/k |
|  |  | 0.090 | -0.020 | 0.070 |  |  |  |
| **Reference values (healthy)** | **10** | **0.275** | **0** | **0.275** | **0.040** | **200** |  |

**Table 2: Summary of the material parameters assigned to the simulation model.**

| component | material model | property |
|---|---|---|
| cornea | anisotropic, hyperelastic, incompressible | $K = 10\ MPa$ |
| | | $\mu = 0.275\ MPa$ |
| | | $k_1 = 0.04\ MPa$ (Healthy) |
| | | $k_1 = 0.02\ MPa$ (KK-I) |
| | | $k_1 = 0.01\ MPa$ (KK-II) |
| | | $k_1 = 0.00\ MPa$ (KK-III) |
| | | $k_2 = 200\ MPa$ |
| limbus | isotropic, linear-elastic, incompressible | $E_{Li} = 1.4\ MPa, \nu = 0.49$ |
| sclera | isotropic, linear-elastic, incompressible | $E_{Sk} = 2.3\ MPa, \nu = 0.49$ |
| lens | isotropic, linear-elastic, incompressible | $E_{Le} = 2.4\ MPa, \nu = 0.49$ |
| aqueous humor | incompressible | $K_W = 2.0$ GPa |

Besides the cornea, the eye model contains further components, which receive reasonably simplified properties, namely linear-elastic, isotropic and incompressible material properties. From the investigations on scleral specimens by means of tensile tests by Friberg and Lace [27], the inflation tests by Woo, Kobayashi, Schlegel and Lawrence [28] and Battaglioli and Kamm [29] as well as from applanation analyses in [30], an averaged elastic modulus of the sclera of $E_{Sc} = 2.3$ MPa is obtained. As a link between cornea and sclera, the limbus obtains an averaged property of both components: $E_{Li} = 0.5(E_{Co} + E_{Sc})$. Here, corneal stiffness refers to the publication [11] about the behavior of the eye shell under IOP. Lens capsule, cortex and nucleus as well as zonular fibers and trabecular meshwork are combined into one body. Its properties correspond to those from the studies of Danielsen [31] on lens capsules. For the aqueous humor, the compression modulus of water is given by $K_W = 2.0$ GPa.

The specified boundary conditions are subdivided into the mounting of the eye, the intraocular pressure IOP and the pressure and shear stresses applied to the corneal surface by the NCT. The displacement and rotation-free positioning of the eye takes place on its posterior surface. All nodes of the scleral outer surface that are within an angle of 30° to the symmetry axis of the eye are fixed. The symmetry-related simplification to a quarter model results in intersecting surfaces. These are provided with floating bearings according to the symmetry constraints. The fluid elements for representing the anterior chamber of the eye and the vitreous body receive a hydrostatic pressure which is fixed at 17.5 mmHg during the determination of the stress-free geometry. Once the geometry is determined and the preload is applied, the transfer of the boundary condition IOP to a degree of freedom is done. This allows the IOP to increase during corneal deformation, as it does in reality. The pressure and shear stresses on the corneal surface caused by the NCT are applied in a location-, time- and deformation-dependent manner. For this purpose, analytical equations are used, which were determined on the basis of experimental and numerical flow investigations and have already been published in [32]. To represent an almost steady development of the load profile, a stepwise simulation of the NCT is performed, i.e. in each load step the changing load profile according to the analytic equation in [32] is applied.

As previously mentioned, the real eye is prestressed during tomographic or topographic in vivo measurements due to the IOP. The stress-free geometry is unknown. An FE model built from these data would have the geometry of the prestressed eye in the stress-free state. When the IOP is applied, the model deforms. This results in a deformed geometry. In order to reduce influences due to these deformations to a minimum, an iterative calculation of the stress-free geometry is performed according to the publication by Pandolfi and Holzapfel [33]. The distance of the nodes between the target geometry and the nodes under prestress is used as a criterion for termination. The criterion is fulfilled if this distance is less than or equal to 2 µm for all nodes.

**Inverse Method for Linear Parameters**

The basis of an FE simulation is the iterative search for equilibrium between the external loads acting on the body and the internal forces in the body derived from the strain energy induced by the body deformation. Equilibrium is achieved when the residuum defined as difference between discrete internal $\boldsymbol{f}^{int}(\hat{\boldsymbol{u}}, \boldsymbol{\alpha})$ and external $\boldsymbol{f}^{ext}$ force matrices approaches zero by iteratively changing the nodal displacement matrix $\hat{\boldsymbol{u}}$. The discrete residual $\boldsymbol{R}$ is calculated as

$$\boldsymbol{R} = \boldsymbol{f}^{int}(\hat{\boldsymbol{u}}, \boldsymbol{\alpha}) - \boldsymbol{f}^{ext} \tag{6}$$

with the material parameter matrix $\boldsymbol{\alpha} = (K, \mu, k_1, k_2)^T$. The matrix with discrete nodal entries of internal forces is obtained by standard assembly procedure of the element-wise column matrix

$$\boldsymbol{f}_e^{int}(\hat{\boldsymbol{u}}, \boldsymbol{\alpha}) = \int_{V_e} \boldsymbol{B}^{eT} \boldsymbol{\sigma}(\hat{\boldsymbol{u}}, \boldsymbol{\alpha}) \, dv = \sum_{i=1}^{3} \alpha_i \underbrace{\int_{V_e} \boldsymbol{B}^{eT} \hat{\boldsymbol{\sigma}}_i(\hat{\boldsymbol{u}}) \, dv}_{\boldsymbol{c}_i} \tag{7}$$

which is identified as integral over the element volume $\boldsymbol{V}_e$ of the standard $\boldsymbol{B}^e$-matrix multiplied by the Cauchy stress matrix $\boldsymbol{\sigma}$ in Voigt notation. The $\boldsymbol{B}^e$-matrix is an element-specific matrix and contains the derivatives of the element ansatz functions [35]. The stresses $\boldsymbol{\sigma}$ are computed as matrix representation of the tensor product $\sigma_{jk} = 2/J \, F_{jm} \, \partial \Psi / \partial C_{mp} \, F_{pk}^T$. In the model proposed by Sánchez, Moutsouris and Pandolfi [14], the parameters $K, \mu$ and $k_1$ enter the strain energy density linearly in three separate additive terms and thus, the internal forces can be split into three additive terms $\alpha_i \boldsymbol{c}_i$ with $i = 1, 2, 3$ and $\alpha_1 = K, \alpha_2 = \mu, \alpha_3 = k_1$. Herein, $\boldsymbol{c}_i$ represents the column matrix as defined in Eq. (7) Thereby, the residuum turns out to be a linear function in these three parameters which is why we refer to these as *linear parameters*. In order to compute these from solving the inverse problem, the nodal displacement matrix $\hat{\boldsymbol{u}}$ is assumed to be fully known here from the NCT measurement. Note that currently the full-field data cannot yet be measured due to technical limitations in the NCT devices. Therefore, we propose to construct an approach which provides a reasonable approximation of the full field kinematics, but this will be addressed later in this paper. Provided that the displacements as well as the external forces are known, the material parameters $\boldsymbol{\alpha}$ remain the only unknowns in the residuum $\boldsymbol{R}$. Assuming that the *nonlinear parameter $k_2$*, which appears nonlinearly in the residuum, is known during the identification of the linear parameters, the residuum can be rewritten as

$$\boldsymbol{R} = K\boldsymbol{c}_\text{K} + \mu\boldsymbol{c}_\mu + k_1\boldsymbol{c}_{\text{k}_1} - \boldsymbol{f}^\text{ext} \tag{8}$$

Herein, the matrix $\boldsymbol{c}_i$, i.e. $\boldsymbol{c}_\text{K} \coloneqq \boldsymbol{c}_1, \boldsymbol{c}_\mu \coloneqq \boldsymbol{c}_2, \boldsymbol{c}_{\text{k}_1} \coloneqq \boldsymbol{c}_3$, contains the result of the integral evaluation and is constant for a given displacement field. The basic idea of the equilibrium gap method (EGM) is that the parameters can be identified by minimizing $\boldsymbol{R}$, with the main advantage that the residuum has only to be evaluated. This is in contrast to updating procedures where the expensive forward problem needs to be solved within each optimization step. The main interpretation of the EGM is that the obtained optimal choice of the parameters is associated with the best possible achievement of equilibrium given a specific displacement field, external forces, and material model. Since neither the material model nor the measurement of displacements or external forces can never be precise, exact equilibrium will never be reached. However, in a post-processing step, the forward problem may be solved using the identified parameters and then the calculated displacement field can be compared with the measured one. If then the similarity is found insufficient, improvements can only be achieved by improving the material model as the EGM already resulted in the optimal values of parameters. In order to exploit the fact that the residuum is linear in the linear parameters, minimization of $\boldsymbol{R}^\text{T}\boldsymbol{R}$ renders a quadratic optimization problem still in line with the original idea of the EGM. The main advantage is that this quadratic and thus convex problem has a unique solution and enables thereby a unique identification of at least the linear parameters. The property that if the derivative of the strain energy function with respect to the deformation tensor is linearly dependent on the material parameters, these can be determined uniquely and directly, has already been observed by Cottin, Felgenhauer and Natke [36]. Due to the fact that the optimization problem is convex, the necessary and sufficient conditions for the optimum are

$$\frac{\delta \boldsymbol{R}^\text{T}\boldsymbol{R}}{\delta K} = 0, \quad \frac{\delta \boldsymbol{R}^\text{T}\boldsymbol{R}}{\delta \mu} = 0 \quad \text{and} \quad \frac{\delta \boldsymbol{R}^\text{T}\boldsymbol{R}}{\delta k_1} = 0 \;. \tag{9}$$

Since $\boldsymbol{R}^\text{T}\boldsymbol{R}$ is quadratic, the derivatives in Eq. (9) will be linear in the parameters $K, \mu, k_1$ and thus, only a linear system of three equations has to be solved to compute the optimal values of linear parameters. A direct consequence is that in principle no iterative numerical optimization procedure is necessary rendering the approach real-time capable. However, usually constraints regarding the parameters need to be considered, for instance, here $K \geq 0, \mu \geq 0, k_1 \geq 0$ due to convexity conditions, and thus, it may happen that the conditions in Eq. (9) would be fulfilled within the precluded parameter regions. Then, an iterative numerical procedure is required to track the parameter space boundaries until the optimum of $\boldsymbol{R}^\text{T}\boldsymbol{R}$ is found. But this case did not appear in our analysis, and even if, the fact that the calculation of the objective function does not include the expensive numerical solution of boundary value problems still renders the numerical procedure almost real-time capable. Note that the implementation using standard Finite Element software for the calculation of the residuum can be obtained in a straightforward, non-invasive manner. Then, for the calculation of the individual matrices $\boldsymbol{c}_i$, simply the associated parameter is set to $\alpha_i = 1$ and the remaining parameters are set to $\alpha_j = 0, j \neq i$. Furthermore, the external forces are set to zero. Then, the FE software is called and the returned discrete residuum matrix represents the matrix $c_i$. For example, if the matrix associated with $K$ has to be computed, the FE software can directly provide $c_K = R\ (K = 1, \mu = 0, k_1 = 0, f^{\text{ext}} = 0)$.

**Inverse Method for the Nonlinear Parameter**

The dimensionless material parameter $k_2$, which was previously considered known, has a nonlinear effect on the transversely isotropic part of the strain energy function in Eq. (5) and thus, on the optimization problem as part of the parameter identification. Accordingly, an unambiguous determination according to the presented approach is not possible. The parameter mainly influences the nonlinearity of the stress-strain curve of the collagen fibers. Consequently, several supporting points of different deformation states are necessary for its determination. In [37] it is proposed to determine such parameters by an additional identification loop. For an initialized and fixed $k_2$, the determination of the linear material parameters $\alpha_i$ is separately performed for $m \geq 3$ different deformation states. Thereby, a set of separate values of linear parameters for each deformation state is obtained. Of course, eventually, a fixated set of parameter values should be obtained no matter which deformation state. Therefore, the similarity between the separate material parameter sets serves as a quality criterion for the selected $k_2$ [37]. To define a suitable criterion, we propose the sum of the standard deviations $\sigma_{\text{Dev}}$ of the material parameter sets related to their mean values $\langle \cdot \rangle$, and thus, an external objective function is defined as

$$f_{\text{rel}}(k_2) = \sum_{i=1}^{3} \frac{\sigma_{\text{Dev},\,\alpha_i}(k_2)}{\langle \alpha_i(k_2) \rangle}. \tag{10}$$

This function depends on $k_2$ since the identification of the separate linear parameter sets depends on the specific choice of $k_2$. By minimizing $f_{\text{rel}}$ with respect to $k_2$, the specific value for the nonlinear parameter is obtained which enables an optimal representation of the degree of nonlinearity of the material response. Note that this overall procedure represents a nested optimization problem because for each evaluation of $f_{\text{rel}}$ several optimization problems for the linear parameters according to the different deformation states need to be solved. Since $f_{\text{rel}}$ is usually not convex in $k_2$, neither a unique identification nor the calculation of the global minimum can be guaranteed. In addition to that, a global non-convex optimizer needs to be applied rendering the external minimization problem comparatively expensive. However, the identification of $k_2$ does not necessarily need to be performed patient-specific. On the contrary, it is rather reasonable to identify this nonlinear parameter first based on experimental data and then, while keeping $k_2$ fixated as part of the model formulation, only the linear parameters are identified patient-specifically in real time.

**Representative Full-Field from Reality-like Data**

The previous approach is based on the knowledge of the nodal displacement matrix $\hat{u}$. Since no measurement system is available so far to measure the full field during air pulse tonometry, the displacement matrix must be approximated otherwise. We propose the following mechanical morphing approach to make use of the time-resolved deformation contours of the anterior and posterior corneal surface determined with the Corvis® ST in such a way that a representative full field is obtained which is sufficiently comparable to the real displacement full field to be able to determine material parameters according to the inverse approach described above. The (i) basic geometry of the eye, the (ii) IOP and the (iii) deformation contours are assumed to be given by the Corvis® ST and enter the approach as input data. From the geometry data, an eye model similar to the one previously explained is created and meshed. Under the assumption that the deformation behavior is essentially characterized by the incompressibility condition and since at this point no information is available regarding the mechanical properties of the cornea belonging to the data, the material is assumed with linear-elastic properties. According to [33], the stress-free geometry is determined from the model under IOP. Then the 2D deformation contours are imported and rotationally extracted to 3D stamps. The stamp of the front deformation contour is indented according to the deflection amplitude $DefA$. Then the stamp of the posterior deformation contour is aligned with the anterior stamp according to the central corneal thickness CCT. The stamps and the cornea are both provided with contact elements on the surface. These enable vertical force transmission so that the new corneal shape is imprinted. At the same time, the cornea can move horizontally by sliding. This is important because the Corvis® ST data only provides the information of the contour, not the displacements of specific points. Thereby, the eye model contour is transferred to the measured contour which is why we refer to this process as mechanical morphing approach. During the process, all model nodes take a displacement assumed to be similar to reality. These can then be read out as a representative displacement matrix $\hat{u}_{\text{morph}}$. The process is shown in Figure 3a and illustrated by Figure 3b.

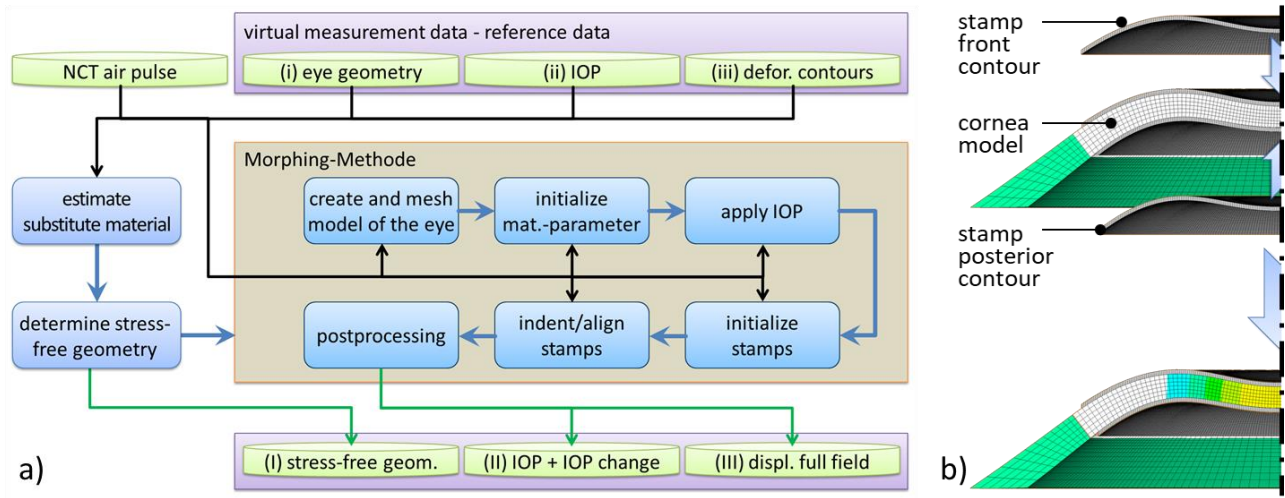

**Figure 3: a) process diagram of the transformation of measurement data (geometry, IOP, deformation contours) into the input data required for the inverse method (stress-free geometry, IOP including the IOP change, displacement full field) using the mechanical morphing approach and b) illustration of the principle of the mechanical morphing approach. Virtual stamps force the corneal model to deform.**

In addition, the deformation of the cornea leads to a pressure increase in the fluid-filled interior of the eye. This IOP change is a non-negligible component of the external forces acting on the cornea and is relevant for the inverse parameter identification. Thus, the mechanical morphing approach has transformed the input variables (i) eye geometry, (ii) IOP and (iii) deformation contours into the output variables (I) stress-free geometry, (II) IOP including the IOP change and (III) the approximated displacement full field $\hat{u}_{\text{morph}}$.

In order to be able to analyze the performance of this approach, the reference displacement matrices already obtained from the NCT simulation are used to construct the according 2D deformation contours as reference, because these would be obtained from the NCT measurement. For each reference data set, one contour for the front and back surface is created and then only this data is used as virtual experimental data instead of the full-field kinematics.

## 3 Results

The eye model is now used for the virtual NCT examination. Similar to the real eye, the model reacts with a corneal deformation to the applied air pulse. Figure 4a compares the typical deformation behavior of a cornea during NCT with the simulation results. The images of the real cornea are based on Corvis® ST measurements on healthy patients, performed in [38]. The comparison shows a high qualitative similarity. Furthermore, the indentation of the cornea leads to an increase in IOP due to the reduction of the anterior chamber volume. This reaction is shown in the diagram Figure 4b. Note the decoupling of the anterior chamber from the vitreous body caused by the lens. Since the lens surface is smaller on the anterior chamber side than on the posterior side, the pressure in the anterior chamber increases much more. This increase counteracts corneal deformation, which would be greater in the absence of IOP response. It follows that consideration of the lens is highly important for NCT simulation. The results of the typical Corvis® ST parameters $DefA$ and $PD$ as well as the eye pressures in the anterior chamber $IOP_{AC}$ and the vitreous body $IOP_{VB}$ are summarized for the time point of maximum indentation in Table 3.

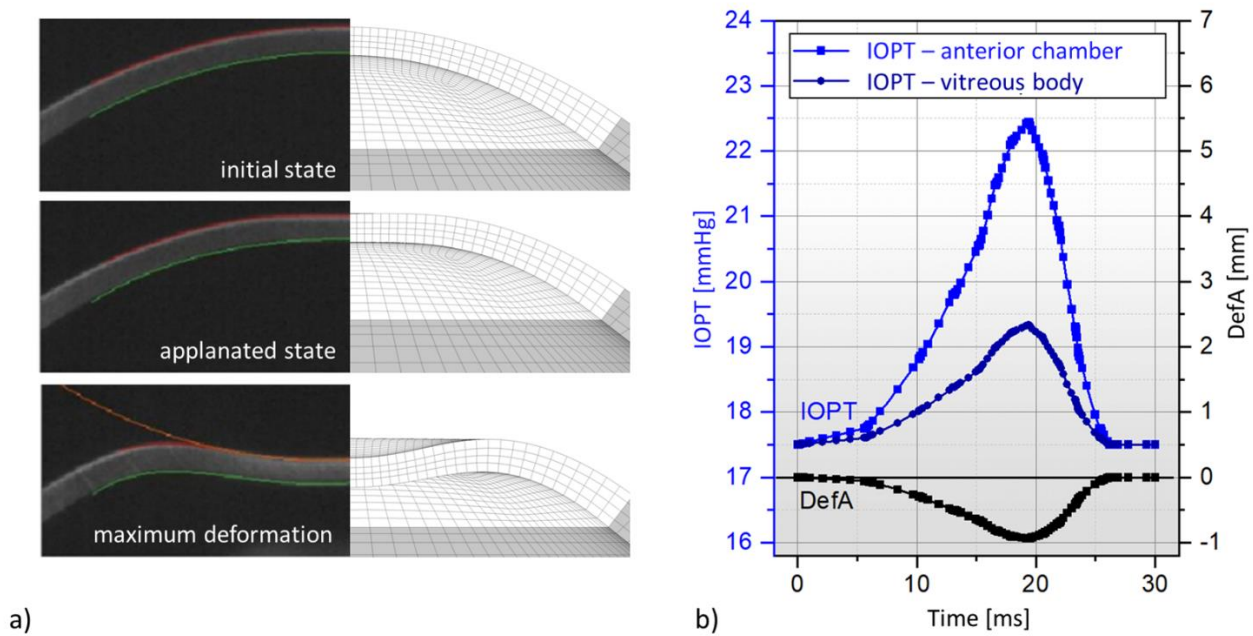

**Figure 4: a) Comparison of the simulated deformation behavior with the deformation states of the cornea recorded by Corvis® ST in sectional view according [38] and b) the calculated profile of the intraocular pressures in the anterior chamber and vitreous body as a function of the deflection amplitude $DefA$.**

**Table 3: Results calculated with the FE model at the time of maximum indentation depth for the 4 predefined material sets.**

| material set | $Def A$ [mm] | $PD$ [mm] | $IOP_{AC}$ [mmHg] | $IOP_{VB}$ [mmHg] |
|---|---|---|---|---|
| Healthy | 0.944 | 4.870 ± 0.108 | 23.264 | 19.564 |
| KK-I | 1.019 | 5.068 ± 0.108 | 23.792 | 19.781 |
| KK-II | 1.058 | 5.061 ± 0.108 | 23.989 | 19.864 |
| KK-III | 1.193 | 5.047 ± 0.108 | 24.412 | 20.052 |

The basis for the verification of the inverse method is the synthetic measurement data, which were extracted from the NCT simulation. In addition to the stress-free geometry, three deformation states are used. State 1 is the applanation state, state 2 defines the deformation state in the middle between applanation and maximum deformation, and state 3 describes the maximum deformation. For these three states, the displacement full-field for each material data set (H, KK-I, KK-II, KK-III) is extracted. This results in 12 reference data sets which can be used to measure the performance of the method.

**Test with Linear Parameters**

The solution of the system of equations (9) with directly inserting the displacement full-field data and the externally acting forces provides the resulting parameters summarized in Table 4. Since this investigation is intended to provide the proof of function for the linear parameters, the nonlinear parameter $k_2$ was kept constant according to the reference value of 200, which was identified in the following subsection. The inverse approach is shown to work very accurately based on the reference data for all three deformation conditions. The identification time was $t_{EGM} \approx 20$ s using 1 core running at 3,2 GHz (Intel i7-8700). In comparison, a single forward solve of the NCT investigation takes $t_{FEM} > 6$ h using 1 core at 3,2 GHz (Intel i7-8700) without previous calculation of stress-free geometry and $t_{FEM} > 11$ h with calculation of the stress-free geometry, respectively. This underlines the advantage of incorporating a non-updating scheme for the inverse problem.

**Table 4: Comparison of the linear material parameters identified inversely for different deformation states to the reference values for a constant $k_2 = 200$.**

| material set | H | | | KK-III | | | reference |
|---|---|---|---|---|---|---|---|
| deformation state | 1 | 2 | 3 | 1 | 2 | 3 | |
| $K$ [MPa] | 9.99997 | 10.00000 | 10.00030 | 9.99997 | 9.99995 | 10.00000 | 10 |
| $\mu$ [MPa] | 0.275010 | 0.274996 | 0.275006 | 0.275008 | 0.275002 | 0.275003 | 0.275 |
| $k_1$ [MPa] | 0.039995 | 0.040000 | 0.039998 | 2.88E-6 | 2.42E-10 | 6.40E-6 | 0.04 \| 0.00 |

**Test with the Nonlinear Parameter**

To identify the nonlinear parameter $k_2$, the linear parameters are determined as a function of the varied parameter $k_2$. The results for the healthy reference data set are shown in the diagram in Figure 5a for the linear parameters $K$, $\mu$ and $k_1$ for the three deformation states. It also shows the nonlinear influence of $k_2$. The curves of the three states intersect at $k_2 = 200$, where apparently the deviations of the deformation-state dependent linear parameter sets are minimal. The diagram Figure 5b shows the behavior of the objective function according to Eq. (10). It shows the intersection point recognizable in Figure 5a as a global minimum for $k_2 = 200$, as well as a local minimum at $k_2 \rightarrow 0$. Quantitatively the global minimum differs clearly from the local one, nevertheless a grid search from near zero upwards seems to be recommended for the secure identification of $k_2$. The linear parameters associated with the identified, optimal value for $k_2$ are presented in Table 4. The entire grid search in the range $10 \leq k_2 \leq 400$ with step size 10 took $t_{\text{EGM,NL}} \approx 720$ s when running the EGM in parallel on 3 cores at 3,2 GHz (Intel i7-8700). The comparable forward simulation would have required $t_{\text{FEM}} \approx 10$ d.

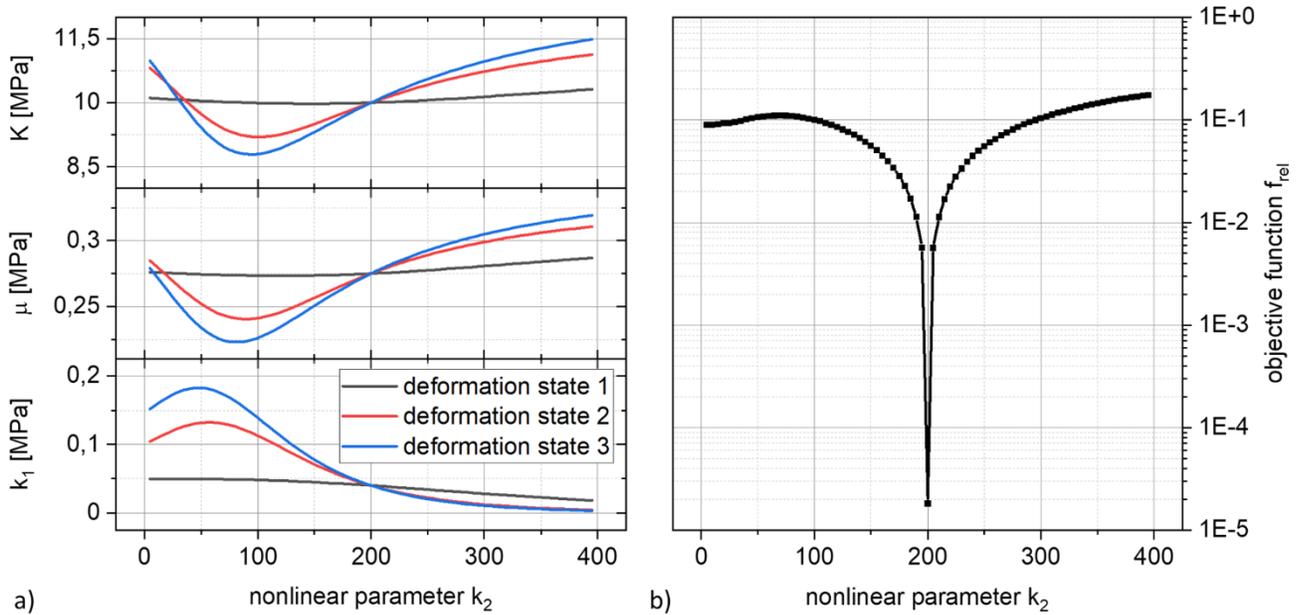

**Figure 5: a) Evolution of the linear material parameters as a function of the nonlinear parameter $k_2$ and b) evolution of the error criterion $f_{\text{rel}}$ according to Eq. (10) over $k_2$.**

The diagram in Figure 6 presents the results of the parameter identification for all 4 reference data sets H, K-I, K-II, and K-III. Similar to the identification of the linear parameters, the results of the identification with nonlinear parameter $k_2$ are again quite accurate. As before, the fiber stiffness parameter $k_1$ follows the predefined degradation. The new nonlinear and dimensionless material parameter $k_2$, which is now included in the identification, is correctly determined for the material parameter sets H, K-I and K-II. For the reference data set K-III, no identification is achieved with reference to the $k_1 \approx 0$. From Eq. (5) it can be concluded that $k_1 \approx 0$ switches off the transversely isotropic part of the strain energy density function, in which $k_2$ is also located. Thus, $k_2$ does not influence the minimization of the residuals anymore.

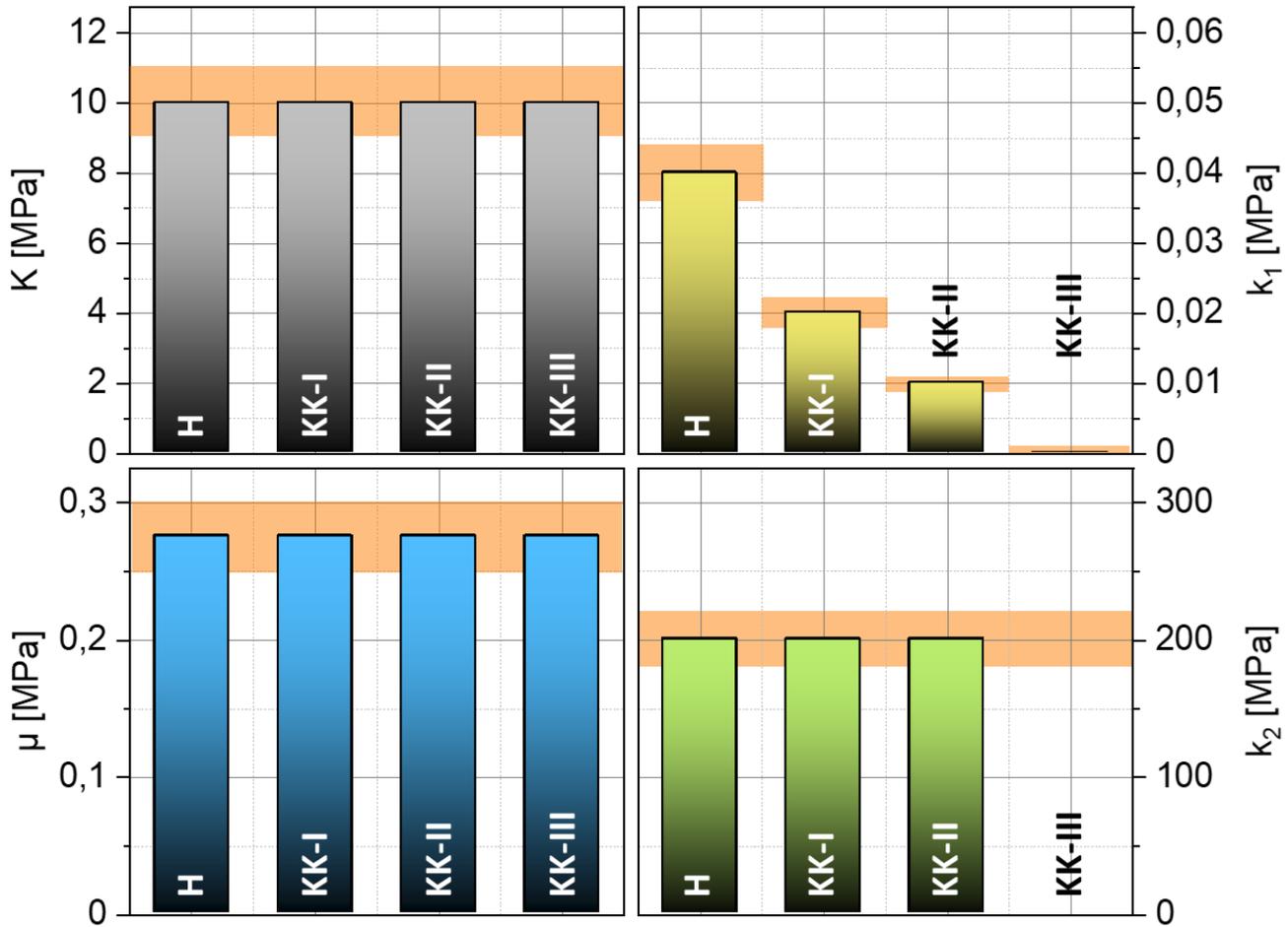

**Figure 6: Inversely identified material parameters of the 4 reference data sets. The diagram shows the ±10% deviation band from the reference value (orange).**

**Influence of Noise**

The previous investigations are based on the use of the unperturbed displacement full field associated to the reference data. However, the advised NCT devices measure the displacement data optically and optical systems are subject to numerous environmental influences that cause image noise. In order to investigate the influence of noisy deformation fields on the inverse parameter identification, a superposition of absolute noise on the displacement full fields of the healthy reference data set (H) is considered. For this purpose, a random offset between $-0.01$ µm to $0.01$ µm or between $-0.1$ µm to $0.1$ µm is added to each component of the displacement matrix $\hat{\boldsymbol{u}}$. For comparison, the nodal displacements in the corneal region have average values of approximately $80$ µm. For random calculation, the RAND function implemented in ANSYS® coupled with the system time is applied. A regularization of the generated noisy displacement matrix $\hat{\boldsymbol{u}}$ in terms of smoothening is intentionally not performed. The diagram Figure 7 compares the results of the inverse parameter identification in the presence of noise. As can be observed, the results are clearly unstable as soon as absolute noise of 0.1 μm is taken into account.

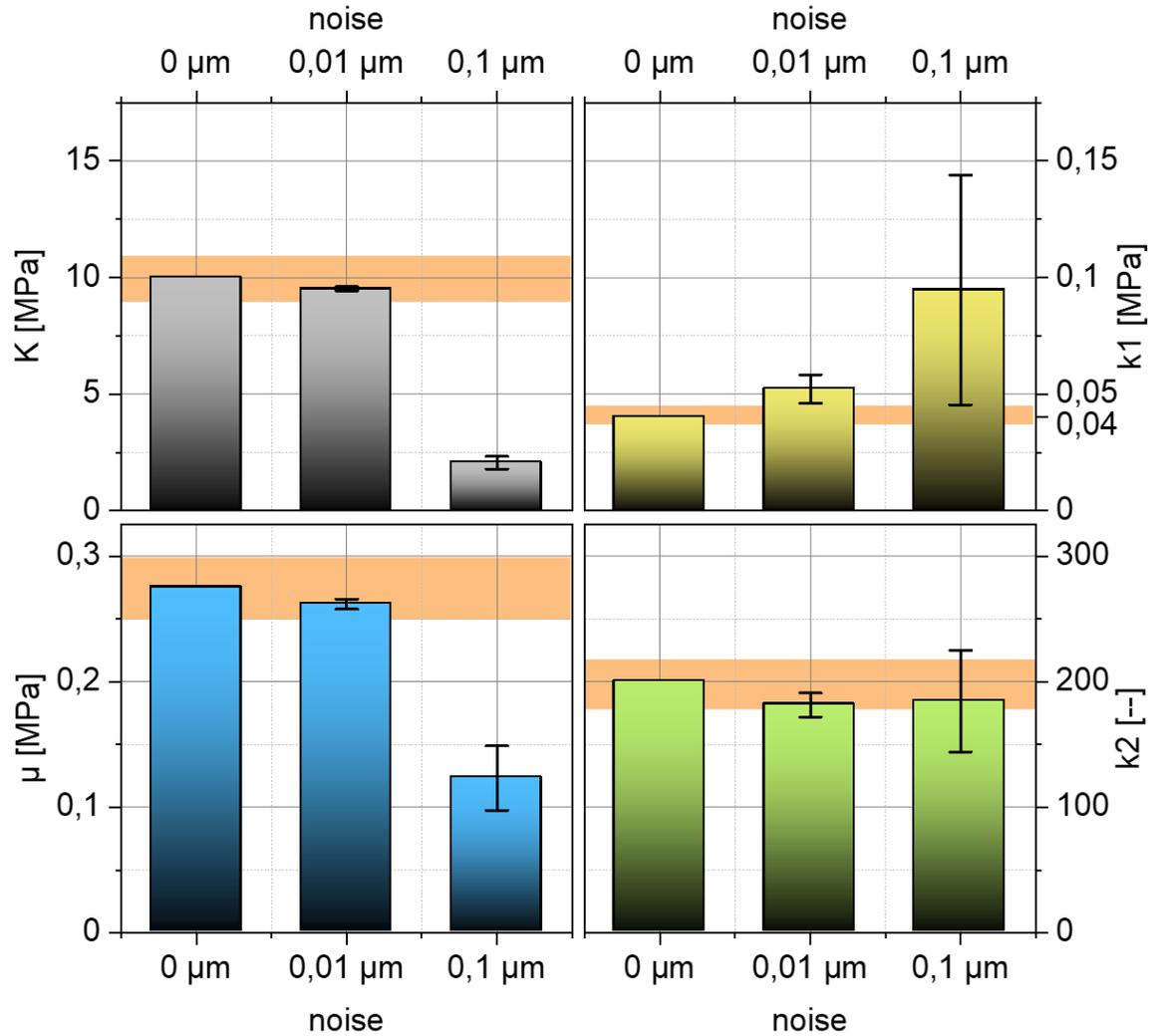

**Figure 7: Inversely identified material parameters from the healthy material set as a function of the applied random noise. The diagram contains the ±10% deviation band to the reference value (orange) as well as the twofold standard deviations determined from 10 repetitions as error bars.**

However, this observation should be put into perspective. Keep in mind that here we considered a rather unrealistically intense noise on purpose in order to challenge the method. The considered noise is random, not Gaussian distributed as may be expected in reality. Furthermore, for the real application unsmoothed, raw pixel data would not be directly considered as geometry input and rather a smoothened surface line would be constructed. In addition to that, the considered noise is completely random in orientation. Considering the 8 nodes of an element, the noise can lead to expansion or contraction of the element volume. The probability that an element contracts or expands because of the noise is thereby many times higher than that the element volume remains constant, since constancy requires all 8 nodes to be shifted by the same offset in the same direction. The violation of incompressibility caused by the noise leads to the underestimation of the penalty parameter $K$. So, the equilibrium between the components of the strain energy function shifts. As a result, the material parameters must compensate the volume change in addition to the deformation and therefore deviate from the target values. This effect increases with the increase of the noise intensity. Interpretation of the noise intensity applied here is rather difficult. Surely, the limitations of the Corvis® ST regarding the resolution of about 15 µm/px (564 px [39] to about 8.5 mm [40, 41]) are non-negligible. However, it may be fair to state that in principle the use of the raw data from the Corvis® ST should not be directly used and rather smoothened data should be considered instead. One strong limitation resulting from the current state of the art regarding technical possibilities should not be omitted here: although current developments are promising with respect to providing technical solutions for the full 3D measurement of displacements, these full-field data are not yet measurable in the eye during NCT.

**Analysis of Morphing Approach for the Approximation of Full Field Displacements**

The necessary input data of a full displacement field for the presented approach is, as mentioned above, not detectable by the Corvis® ST today. Accordingly, a mechanical morphing approach for the data enrichment was presented, which serves as a link between possible measurement data and the presented inverse identification apparatus. In the following, it will be shown that this approximative full field contains all necessary information to identify the material parameters of the more complex material model according to Sánchez, Moutsouris and Pandolfi. For this analysis, the reference data sets consisting of virtually measured displacement fields have to be reduced to data currently available during NCT. This means to the coordinates of the 2D deformation contours in the intersection from anterior and posterior corneal surface with a symmetry plane of the eye. The diagram in Figure 8 presents graphically the results of the parameter identification combined with the mechanical morphing approach for all 4 reduced reference data sets.

As can be seen, the linear-elastic material model in the approach for estimating an equivalent full field provokes the incompressibility much stronger than the penalty parameter of $K = 10$ MPa given in the reference data set. Since the incompressibility is already ensured with $K = 10$ MPa, higher values do not have much influence on the results of the material parameters. The parameter $\mu$ shows a slightly increasing deviation with increasing degradation. The average of the deviation is about -25 %. The determined $\mu$ is generally smaller than the expected value of 0.275 MPa. The dimensionless material parameter $k_2$ is underestimated for all four material data sets. Its relative deviation is about -60 %. The determined parameter is constant, as specified in the reference data. The relative deviation of the fiber stiffness parameter $k_1$ increases with increasing degradation, as observed for the parameter $\mu$. On average, the deviation is about 30 %. However, the major result here is that despite the overall only moderate accuracy to determine the precise values of the parameters, a collagen stiffness degradation induced by disease was clearly detectable through the identified values of the parameter $k_1$ rendering the proposed approach feasible as diagnostic tool based on NCT measurements.

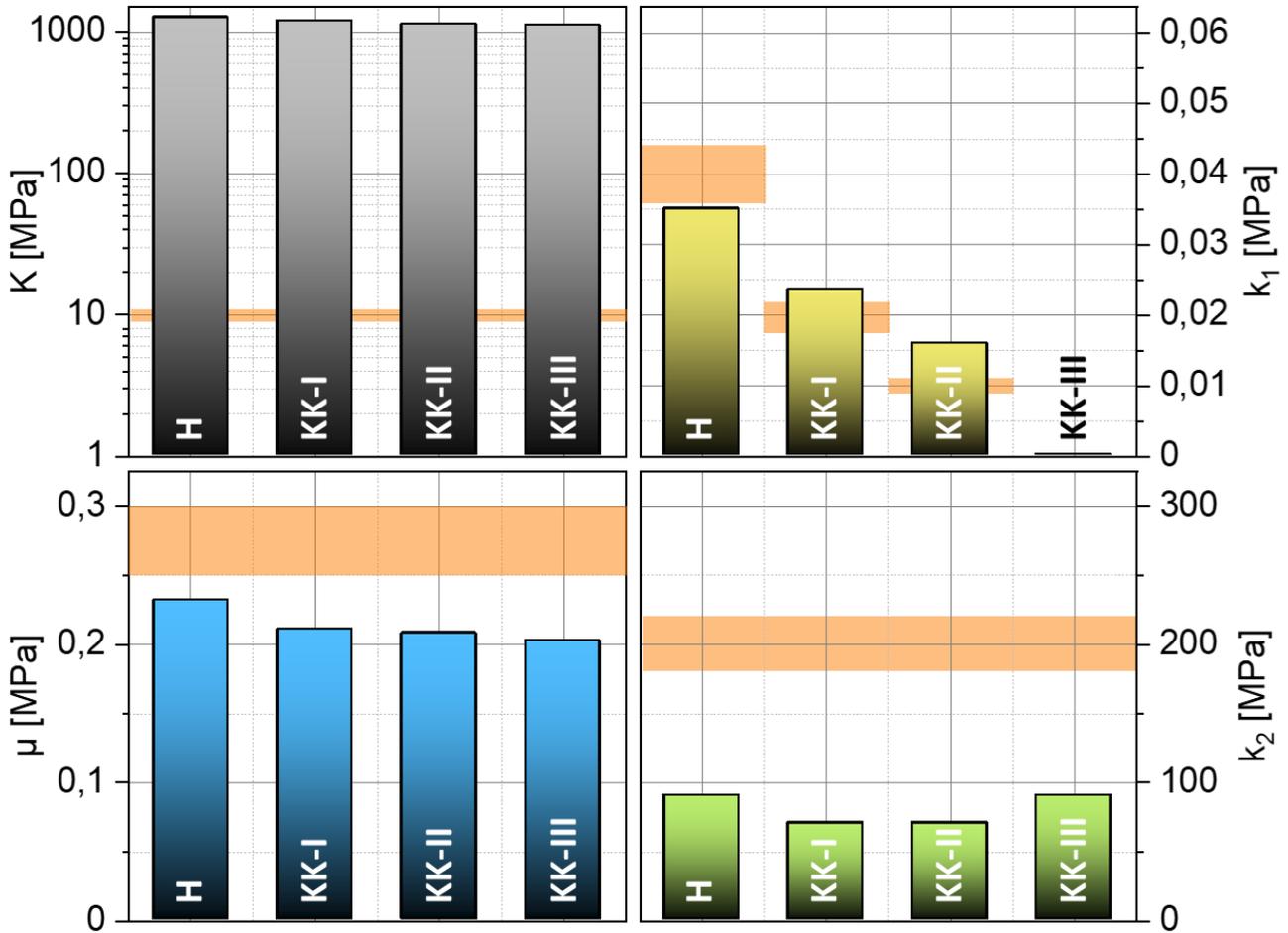

**Figure 8: inverse determined material parameters for the four test cases H, KK-I, KK-II and KK-III as well as the ±10% deviation band to the reference value (orange).**

## 4 Conclusions

The authors proposed a method that enables the in vivo identification of structural material properties of the human cornea while being update-free and thus, furthermore enabling a unique parameter identification in almost real-time. The method is based on the EGM and the approach for the identification of nonlinear parameters in Perotti et al. [37], combined with a new approach to construct approximate full-field kinematics from standard NCT measurements. The functionality of the method was demonstrated on synthetic data of NCT measurements serving as reference.

A finite element model of the human eye was used to synthetically generate displacement full-fields from 4 material parameter datasets with keratoconus-like degradations. Thus, the authors generated reference data sets consisting of virtually measured displacement fields related to prescribed material parameters. This allowed to quantitatively demonstrate the accuracy and potential of the proposed approach of EGM combined with the approach for the identification of nonlinear parameters. In a further step, random absolute noise was added to the displacement fields to investigate the sensitivity to noise. To close the current gap between recordable data and available real-time inverse approach, we proposed and tested an approach for the construction of approximative displacement fields just based on data currently available during NCT, which automatically contains to some extent data smoothing. The authors then quantified the accuracy of the whole method based on the EGM, the nonlinear parameter identification approach, and the mechanical morphing approach.

The inversely identified material parameters based on the known full fields have shown the high accuracy of the approach. In this context, it should be pointed out that there exists currently no technical equipment able to measure the required full fields during air pulse tonometry. This point is supported by the performed noise analysis. It has been shown that the approach is sensitive to random noise for already small amplitudes of 0.1 µm. With an identification time of about 720 s, the approach is much faster than inverse problems solved by typical forward approaches. These would require many thousands of forward simulations of the NCT where each of those would already require more than 11 h when using equivalent computing resources. The results of the identification based on the data currently available during the NCT and using the mechanical morphing approach showed that a degradation of the parameter $k_1$ is detectable. This parameter is related to the stiffness of collagen fibers. The deviations in the parameter identification of about 25 % to 30 % for $\mu$ and $k_1$ and of about 60 % for $k_2$ are too large for subclinical diagnosis. However, analysis show that the accuracy is sufficient for the identification of diseased tissue properties.

An open issue is still the lack of technical equipment to accurately measure the required displacement full fields during the very fast NCT examination. The mechanical morphing approach to overcome this lack was based on a strong simplification, namely linear elasticity. This was motivated by the fact that initially there is no information regarding the material properties and thus, it should be avoided to include too much a priori knowledge. However, further improvements may be achieved by extending accuracy here. Furthermore, our approach assumes knowledge about the stress-free geometry of the eye. Currently, this is not measurable. The procedure of Pandolfi and Holzapfel [33] would be one possible approach to obtain the stress-free geometry. However, it is based on time-consuming forward simulations, which take about 5 h for the presented model. Therefore, new methods and approaches must be investigated in the future.

Finally, the authors highlighted that the proposed framework is able to identify the material properties of human eye tissues in vivo and contactless based on constructed displacement full-fields. During the NCT, the patient's eye is not injured and there is no need to obtain specimens that could be altered by the removal or environmental conditions. Based on data gathered during NCT, our approach enabled the unique identification of degenerated fiber stiffness. Thereby, it represents a promising new method to be used as additional medical diagnostic tool.

**Acknowledgments**

The authors appreciate financial funding from the European Research Fund (ESF) through the training research group "CoSiMa" (ESF Project 100231947) and the "EyeTwin"-Project (ESF Project 100379341).